# Sulfate Radical Oxidation of Aromatic Contaminants: A Detailed Assessment of Density Functional Theory and High-Level Quantum Chemical Methods

Sangavi Pari,[1] Inger A. Wang,[1,a] Haizhou Liu[1]* and Bryan M. Wong[1,2]*

[1] Department of Chemical & Environmental Engineering, University of California-Riverside, Riverside, California, USA

[2] Materials Science & Engineering Program, University of California-Riverside, Riverside, California, USA

[a] Currently at Pennsylvania State University, University Park, Pennsylvania, USA

* Corresponding authors: Haizhou Liu: haizhou@engr.ucr.edu, tel: (951) 827-2076, fax (951) 827-5696
and Bryan M. Wong: bryan.wong@ucr.edu, tel: (951) 827-2163, fax (951) 827-5696, website: http://www.bmwong-group.com

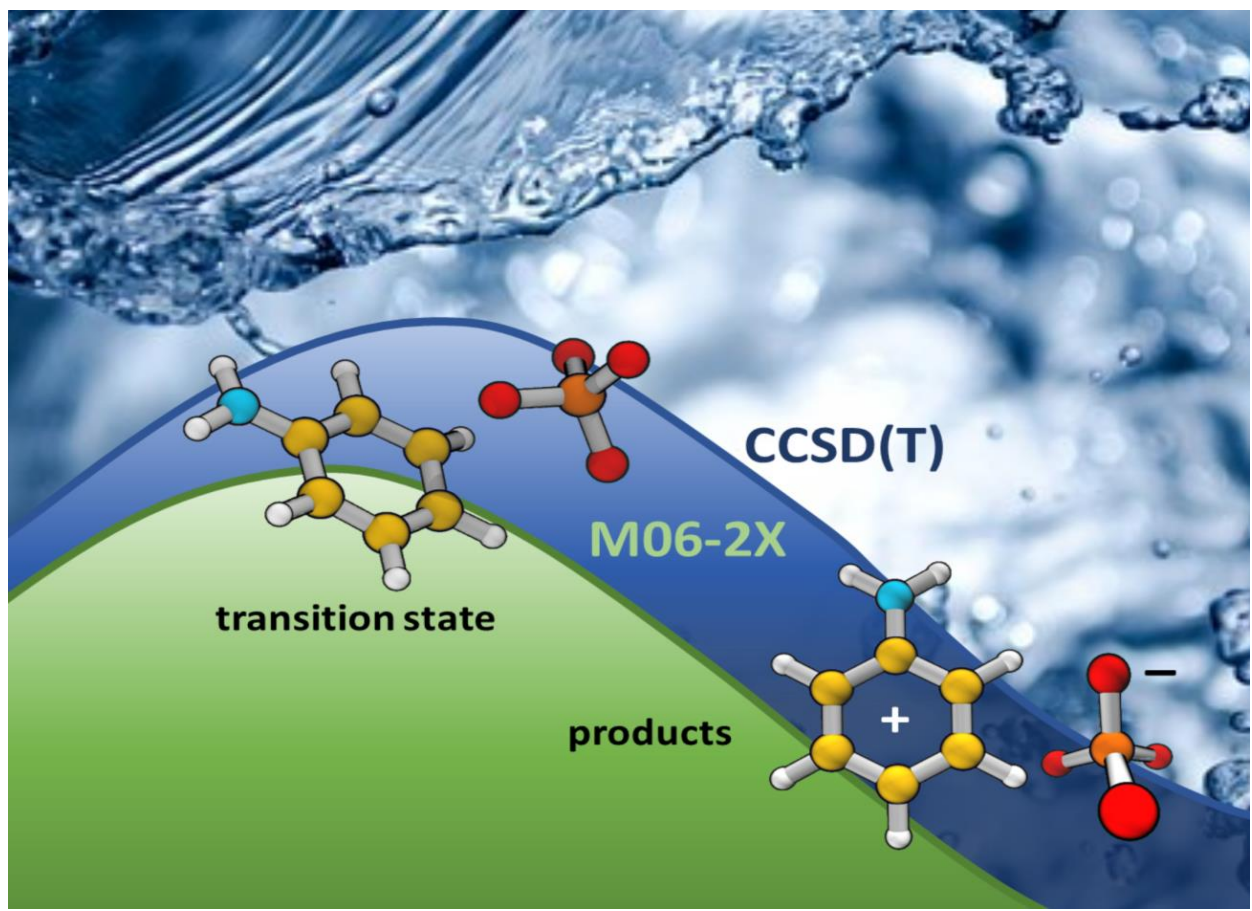

**TOC Figure**

**Abstract**

Advanced oxidation processes that utilize highly oxidative radicals are widely used in water reuse treatment. In recent years, the application of sulfate radical ($SO_4^{\bullet-}$) as a promising oxidant for water treatment has gained increasing attention. To understand the efficiency of $SO_4^{\bullet-}$ in the degradation of organic contaminants in wastewater effluent, it is important to be able to predict the reaction kinetics of various $SO_4^{\bullet-}$-driven oxidation reactions. In this study, we utilize density functional theory (DFT) and high-level wavefunction-based methods (including computationally-intensive coupled cluster methods), to explore the activation energies and kinetic rates of $SO_4^{\bullet-}$-driven oxidation reactions on a series of benzene-derived contaminants. These high-level calculations encompassed a wide set of reactions including 110 forward/reverse reactions and 5 different computational methods in total. Based on the high-level coupled-cluster quantum calculations, we find that the popular M06-2X DFT functional is significantly more accurate for $HO^-$ additions than for $SO_4^{\bullet-}$ reactions. Most importantly, we highlight some of the limitations and deficiencies of other computational methods, and we recommend the use of high-level quantum calculations to spot-check environmental chemistry reactions that may lie outside the training set of the M06-2X functional, particularly for water oxidation reactions that involve $SO_4^{\bullet-}$ and other inorganic species.

**Introduction**

Water scarcity has become a global crisis. This situation is exacerbated – and will continue to be dominated – by the global shrinkage of surface water sources, notably sharp decreases caused by extreme climate conditions.[1, 2] Municipal wastewater reuse offers the potential to significantly increase the nation's total available water resources. Approximately 12 billion gallons of municipal wastewater effluent are discharged each day in the U.S., which is equivalent to 27% of the total public water supply.[3] However, only about 10% of the wastewater effluent is actively reused nationwide.[3] One major challenge to recycling is the development of efficient and cost-effective purification processes. Wastewater effluent is widely compromised by sewage produced from growing populations, industries and agriculture. Trace organic chemicals including petroleum hydrocarbons, pharmaceuticals, personal care products, and industrial solvents are often present in the effluent.[4-13]

To minimize the presence of trace organic chemicals, different advanced oxidation processes (AOPs) have been employed.[14-19] The most widely applied approach is based on the photolysis of hydrogen peroxide ($H_2O_2$) to produce hydroxyl radical ($HO^\bullet$). In recent years, sulfate radical ($SO_4^{\bullet-}$) has garnered much attention as an alternative oxidant for AOP.[20-22] In these processes, $SO_4^{\bullet-}$ is typically generated via UV photolysis of persulfate ($S_2O_8^{2-}$) for water reuse applications.[23-26] $SO_4^{\bullet-}$ has a similar oxidizing power to $HO^\bullet$, yet possessing selectively higher reaction rates with electron-rich contaminants that are typically observed in wastewater effluent.[27-30] Due to a higher quantum yield, the rate of $S_2O_8^{2-}$ photolysis is 40% higher than that of $H_2O_2$ under UV irradiation at 254 nm (a typical wavelength used in UV lamps).[31] Furthermore, the scavenging effect of $S_2O_8^{2-}$ on $SO_4^{\bullet-}$ is two orders of magnitude lower than the scavenging effect of $H_2O_2$ on $HO^\bullet$,[32, 33] which leads to a higher yield of $SO_4^{\bullet-}$ from $S_2O_8^{2-}$ than that of $HO^\bullet$ from $H_2O_2$. These chemical features make $SO_4^{\bullet-}$-based AOP an attractive option for water reuse.

Considering these prospective applications of $SO_4^{\bullet-}$-based oxidation processes for water reuse, it is important to predict the reaction kinetics of $SO_4^{\bullet-}$-driven oxidation reactions with organic contaminants that are present in wastewater effluent. Although some of the radical-driven rate constants can be measured using experimental techniques, *e.g.*, electron pulse radiolysis and γ radiation,[34, 35] it is logistically unrealistic to experimentally measure the rates of every contaminant with short-lived radical species. In addition, the activation energies of $SO_4^{\bullet-}$ with different benzene-derived contaminants are largely unknown, and understanding the activation energies in possible degradation pathways of organic contaminants on a fundamental level are required for predicting byproduct formation in $SO_4^{\bullet-}$ based oxidative water treatment.

Recently, density functional theory (DFT) calculations have started to become commonplace as computational tools for predicting reaction mechanisms and activation energies in redox reactions of environmental significance. There has been recent work using quantum chemical techniques to estimate reaction barriers and thermodynamic relations in the degradation of trace organic contaminants, especially via the oxidation of $HO^\bullet$ or ozone ($O_3$).[36-41] Due to the complexity of the chemical species examined, these prior studies largely used popular computationally-efficient DFT methods to systematically explore and assess the reactivities of contaminants in the aqueous phase. However, most DFT methods are heavily parameterized against a training set of benchmark molecules, *i.e.*, typically organic compounds containing only hydrogen, carbon, nitrogen, and oxygen.[42-44] Specifically, these DFT methods were developed to minimize errors on a given training set of molecules; however, for systems and properties *outside* the training set, Burke and co-workers have demonstrated that these extrapolations can be prone to large and unpredictable deviations.[45] As a result, additional high-level wavefunction based methods are essential to assess whether popular DFT methods are sufficiently accurate for modeling activation energies and thermochemistry, particularly for *inorganic* species outside typical DFT training sets, in environmental computational studies.

The purpose of this study is to assess the accuracy of popular DFT and high-level wavefunction-based methods in quantifying the activation energies of benzene-derived contaminants reacting with $SO_4^{\bullet-}$. Fig. 1 depicts the chemical structures of the various benzene-derived organic contaminants, and Fig. 2 shows the two steps of oxidation reaction investigated in this work. Step 1 involves the addition of $SO_4^{\bullet-}$ to form a benzene-derivative cationic radical and the $SO_4^{2-}$ anion. The addition of hydroxide $OH^-$ to the benzene-derivative cation in Step 2 gives the final hydroxylated oxidation product. We examined a wide set of reactions (110 forward and reverse reactions, in total) using a variety of computational techniques including DFT, MP2/MP4 perturbation theory methods, and high-level coupled cluster CCSD/CCSD(T) approaches. Statistical analyses were carried out for all of these reactions to assess the strengths and limitations of each of the computational methods. We concluded with a discussion and assessment of the specific methods that provide the best accuracy in describing these specific reaction processes relevant to $SO_4^{\bullet-}$ based oxidative water treatments.

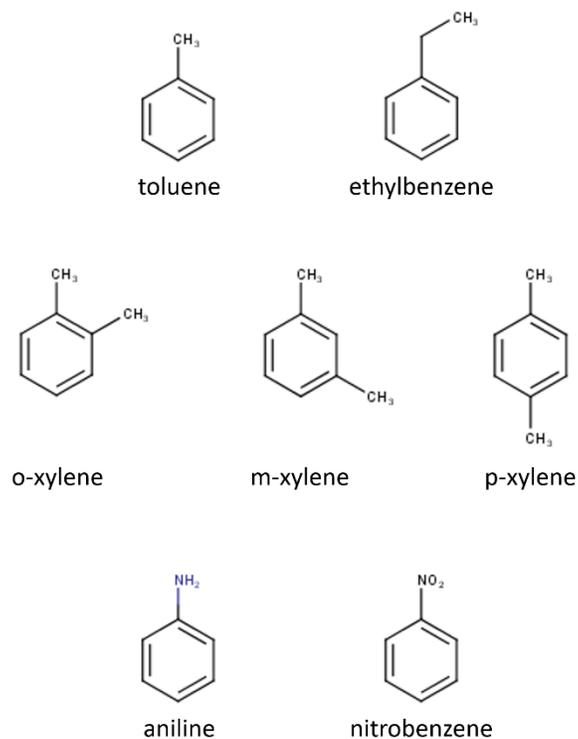

**Figure 1**. Chemical structures of benzene-derived organic contaminants investigated in this work.

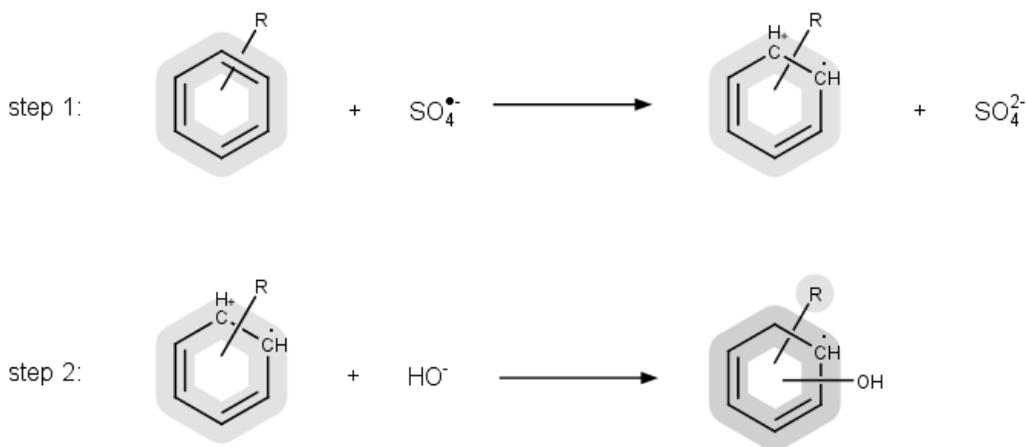

**Figure 2**. Reaction steps investigated in this work: Step 1 involves the addition of $SO_4^{•-}$ to form a benzene-derivative cation and the $SO_4^{2-}$ anion. The *R* group represents the different functional groups depicted previously in Fig. 1. Step 1 involves the addition of HO⁻ to the benzene-derivative cation that gives the final hydroxylated product.

**Computational Methods**

For all of the chemical species investigated in this work, we benchmarked the performance of the highly-parametrized M06-2X DFT functional[46] against the wavefunction-based MP2,[47, 48] MP4(SDQ),[49, 50] CCSD,[51, 52] and CCSD(T)[53] methods. The M06-2X exchange-correlation functional[46] includes 54% Hartree-Fock exchange and has been utilized to study a broad spectrum of chemical species and reactions.[54-59] The MP2[47, 48] and MP4(SDQ)[49, 50] wavefunction-based methods incorporate a Møller-Plesset correlation correction[60] to the total energy based on perturbation theory truncated at second order and fourth order for MP2[47, 48] and MP4(SDQ),[49, 50] respectively. The CCSD[61, 62] and CCSD(T)[53] methods utilize a coupled cluster approach including both single and double excitations (and triple excitations for the CCSD(T) method) to obtain highly accurate total energies. The higher-level CCSD and CCSD(T) theoretical approaches typically provide extremely accurate energies and reaction rates, albeit at a high computational cost. In order to maintain a consistent comparison across the M06-2X, MP2, MP4(SDQ), CCSD, and CCSD(T) levels of theory, the same ground-state and transition-state geometries for all methods were used. Both the ground-state and transition-state geometries were obtained optimized at the M06-2X/6-311+G(d,p) level of theory, and all transition states were confirmed to be first-order saddle points by verifying the presence of a single imaginary harmonic frequency. It is worth mentioning that a complete characterization of a transition state geometry requires a full analysis of the intrinsic reaction coordinate (IRC); however, due to the large number of reactions considered in this study (110 forward/reverse reactions) and the immense computational expense of IRC calculations, we only characterized these transition states with a frequency analysis and reserved the more complete IRC calculations for a future study. For all of the chemical species and computational methods in this study, we utilized the conductor-like polarizable continuum model (PCM) devised by Tomasi and co-workers[63-67] which creates a solute cavity via a set of overlapping spheres to calculate the solvent reaction field.

For all of the wavefunction-based methods (MP2, MP4(SDQ), CCSD, and CCSD(T)) in this study, the same 6-311+G(d,p) basis was also used to calculate total energies for both the ground- and transition-state geometries. Throughout this work we used the CCSD(T) energies as reference values to assess the quality for all of the various methods. We have previously found that the CCSD(T) method accurately reproduces experimental activation energies[68, 69] and electronic properties[70] of various hydrocarbons. As an additional verification on the quality of the CCSD(T) benchmarks, we checked for possible deficiencies inherent to the single-reference coupled cluster approach. Specifically, for open-shell systems, Schaefer and co-workers[71] proposed an open-shell "T1 diagnostic" to determine whether the single-reference-based CCSD procedure is appropriate or requires a higher-level multi-reference treatment. Based on their criterion, if the Euclidean norm of the t1 vector from an open-shell CCSD calculation is greater than 0.044, a higher-level multireference method is necessary. We have computed the T1 diagnostic for all of the geometry-optimized chemical species in this work and found that none of the systems in this study required a multi-reference treatment of electron correlation (open-shell T1 diagnostic values were in the 0.026 – 0.039 range), indicating that all of the chemical species in this study are accurately described by the coupled-cluster approach. All calculations were carried out with the Gaussian 09 package.[72]

**Results and Discussion**

Before proceeding to a detailed discussion of activation energies for the various reactions, we first carried out a series of two high-level benchmark calculations to assess (1) the robustness of the M06-2X optimized geometries and (2) the accuracy of the 6-311+G(d,p) basis set. Due to the computational complexity of these benchmarks, we only performed these calculations on the 1-aminophenol transition-state and final product, as shown in Fig. 3 (as a side note, these benchmark calculations were extremely computationally intensive, with the largest of these calculations taking up to *6 continuous days* on 16 × 2.3 GHz AMD Opteron CPUs and *over 230 GB of disk space* on rapid-access solid state drive storage). To assess the robustness of the M06-2X optimized geometries, we calculated CCSD single-point energies on top of CCSD and M06-2X optimized geometries for the 1-aminophenol transition-state and final product. Fig. 3(a) shows that the difference in CCSD single-point energies obtained from the CCSD and M06-2X optimized geometries is negligible (less than 0.2 kcal/mol), indicating that the M06-2X geometries used throughout this work are reliable. With the M06-2X optimized geometries verified, we next assessed the accuracy of the 6-311+G(d,p) basis set by comparing CCSD(T)/aug-cc-pvtz and CCSD(T)/6-311+G(d,p) single-point energies on top of the same M06-2X optimized geometries used in Fig. 3(a). Fig. 3(b) shows that the difference between the correlation-consistent aug-cc-pvtz and the smaller 6-311+G(d,p) basis set is also relatively small (less than 0.65 kcal/mol), indicating that the 6-311+G(d,p) basis set can be safely used for calculating the thermochemical properties for the numerous reactions (110 forward/reverse reactions) evaluated in this computational study.

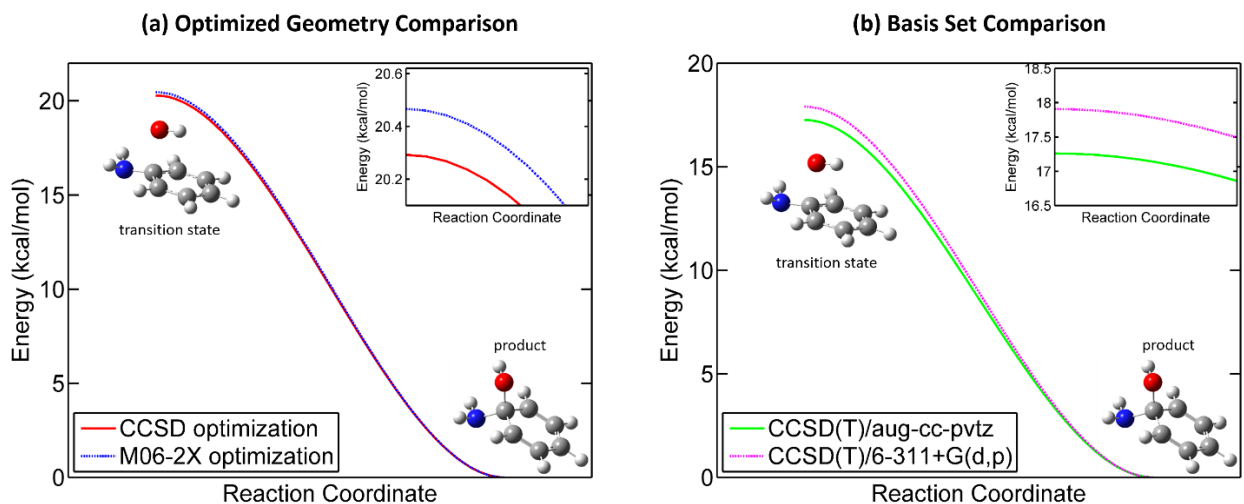

**Figure 3**. Comparison of (a) CCSD single-point energies on top of CCSD and M06-2X optimized geometries and (b) CCSD(T)/aug-cc-pvtz and CCSD(T)/6-311+G(d,p) single-point energies on top of M06-2X optimized geometries for 1-aminophenol. The insets in each of the figures show a magnified portion of the potential energy surface near the transition state, indicating a small energy difference among the various computational methods used in each of the figures.

With these benchmark tests validating our computational approach, we then examined the activation energies for the initial reaction step involving the addition of $SO_4^{\bullet-}$ to the various benzene-derived contaminants. For each of the contaminant, there are multiple sites on the benzene ring that the $SO_4^{\bullet-}$ radical can attach to in the transition state structure. As an example, Fig. 4 depicts the various transition state structures involved in the following reaction: toluene + $SO_4^{\bullet-}$ → toluene cation + $SO_4^{2-}$. We explored all of these possible transition-state geometries for toluene as well as for *all* the chemical species (25 transition states in total) depicted in Fig. 1. Using the CCSD(T) activation energies, $E_a$, as benchmarks, we performed a mean absolute error (MAE) analysis for both the forward and reverse activation energies involving $SO_4^{\bullet-}$ and the various chemical species, which is summarized in Table 1. Fig. 5 depicts in more detail the general trends in the forward and reverse activation energies between the various quantum chemical methods. The diagonal line in all of these figures represents an ideal 100% agreement between the CCSD(T) energies and the other computational methods. It is important to mention that the $R^2$ values listed in Table 1 were obtained from a simple linear fit to the data points themselves and not calculated with respect to the diagonal lines shown in Fig. 5. From the results shown in Fig. 5, it is worth mentioning that the calculated forward activation energy in step 1 is larger than its reverse activation energy; however, the product of the forward reaction in step 1 (i.e., the benzene cation radical) can further react with HO⁻ via the forward reaction in step 2 (Fig. 2). As discussed further in the paragraphs below, the forward activation energy in step 2 is lower than reverse activation energy in step 1. As a result, the rate-limiting step is the forward reaction involving a benzene-derived compound and $SO_4^{•-}$, and once this energy barrier is overcome, the benzene-derivative cation cation radical will further react with HO- to generate the hydroxylated product.

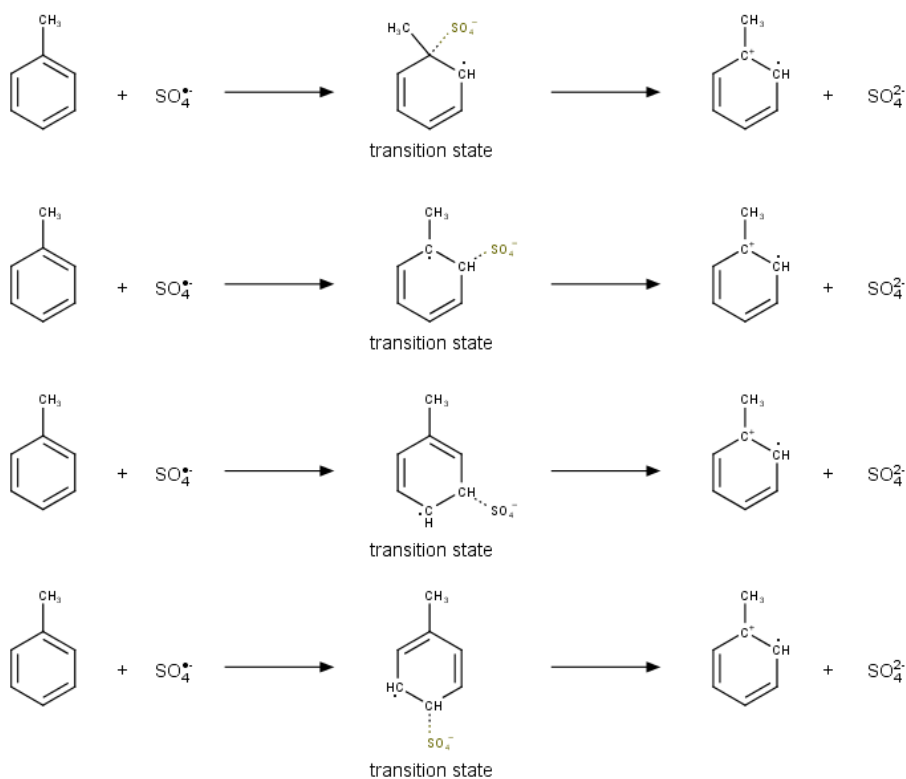

**Figure 4**. Reactants, transition states, and products for the first reaction step involving addition of $SO_4^{•-}$ to toluene.

**Table 1.** Mean absolute error (MAE) with respect to CCSD(T) benchmarks and $R^2$ fit values for various activation energies ($E_a$) computed with the M06-2X, MP2, MP4(SDQ), and CCSD methods.

| | M06-2X | | MP2 | | MP4(SDQ) | | CCSD | |
|---|---|---|---|---|---|---|---|---|
| | MAE (kcal/mol) | $R^2$ | MAE (kcal/mol) | $R^2$ | MAE (kcal/mol) | $R^2$ | MAE (kcal/mol) | $R^2$ |
| **Forward $E_a$ for $SO_4^{\bullet-}$ addition** | 3.35 | 0.95 | 19.34 | 0.69 | 8.23 | 0.83 | 1.02 | 0.99 |
| **Reverse $E_a$ for $SO_4^{\bullet-}$ addition** | 2.00 | 0.99 | 15.37 | 0.38 | 10.11 | 0.86 | 4.20 | 1.00 |
| **Forward $E_a$ for $HO^-$ addition** | 0.88 | 0.99 | 17.51 | 0.83 | 9.56 | 0.96 | 1.69 | 1.00 |
| **Reverse $E_a$ for $HO^-$ addition** | 0.49 | 0.88 | 8.26 | 0.81 | 6.34 | 0.91 | 2.27 | 0.99 |

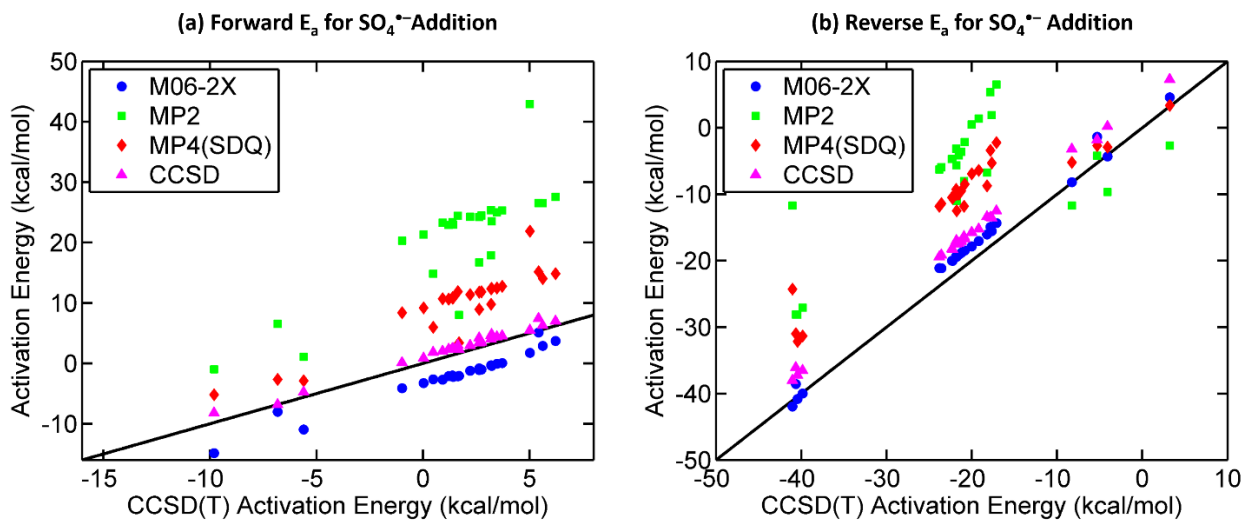

**Figure 5.** Predicted activation energies of the (a) forward and (b) reverse reactions of various benzene derivatives reacting with $SO_4^{\bullet-}$. The diagonal line in each figure represents a perfect match to the benchmark CCSD(T) activation energies.

Both Table 1 and Fig. 5 show that the CCSD calculations (MAE = 1.02 kcal/mol) are in excellent agreement with the CCSD(T) benchmarks for the forward activation energy for the $SO_4^{\bullet-}$ addition. The M06-2X calculations have errors that are quite higher (MAE = 3.35 kcal/mol), followed by the MP4(SDQ) and MP2 methods which have even larger MAEs of 8.23 and 19.34 kcal/mol, respectively. Upon examining the reverse activation energy for the $SO_4^{\bullet-}$ addition, we surprisingly find that the M06-2X functional significantly outperforms the CCSD method by a factor of 2 with respect to the total MAE. Again, the MP4(SDQ) and MP2 methods incur larger errors compared to either the CCSD or M06-2X calculations for the reverse activation energy for the $SO_4^{\bullet-}$ addition. We also note that the CCSD and M06-2X calculations exhibit a high degree of statistical correlation ($R^2 = 0.95 - 1.00$) for both the forward and reverse activation energy for the $SO_4^{\bullet-}$ addition, indicating that the errors in each of these computational methods is systematic rather than random.

We next examined the activation energies for the second reaction step involving the addition of $HO^-$ to the various benzene-derived cation radicals. As before, for each of the benzene cation radical, there are multiple sites on the benzene ring that the $HO^-$ molecule can attach to. As a particular example, Fig. 6 depicts the various products involved in the addition of $HO^-$ to the toluene cation. Again, we explored all of these possible transition-state geometries for toluene as well as for *all* the chemical species (30 transition states in total) depicted in Fig. 1, yielding the various products shown in Fig. 6. A mean absolute error (MAE) analysis was carried out for both the forward and reverse activation energies for all of these resulting products using the CCSD(T) activation energies as benchmarks. Table 1 and Fig. 7 summarize and depict the general trends in the forward and reverse activation energies between the various quantum chemical methods.

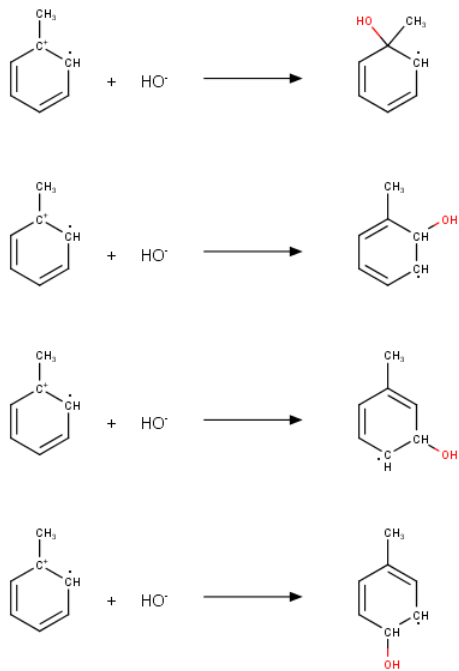

**Figure 6**. Reactants and products for the second reaction step involving addition of HO$^-$ to toluene.

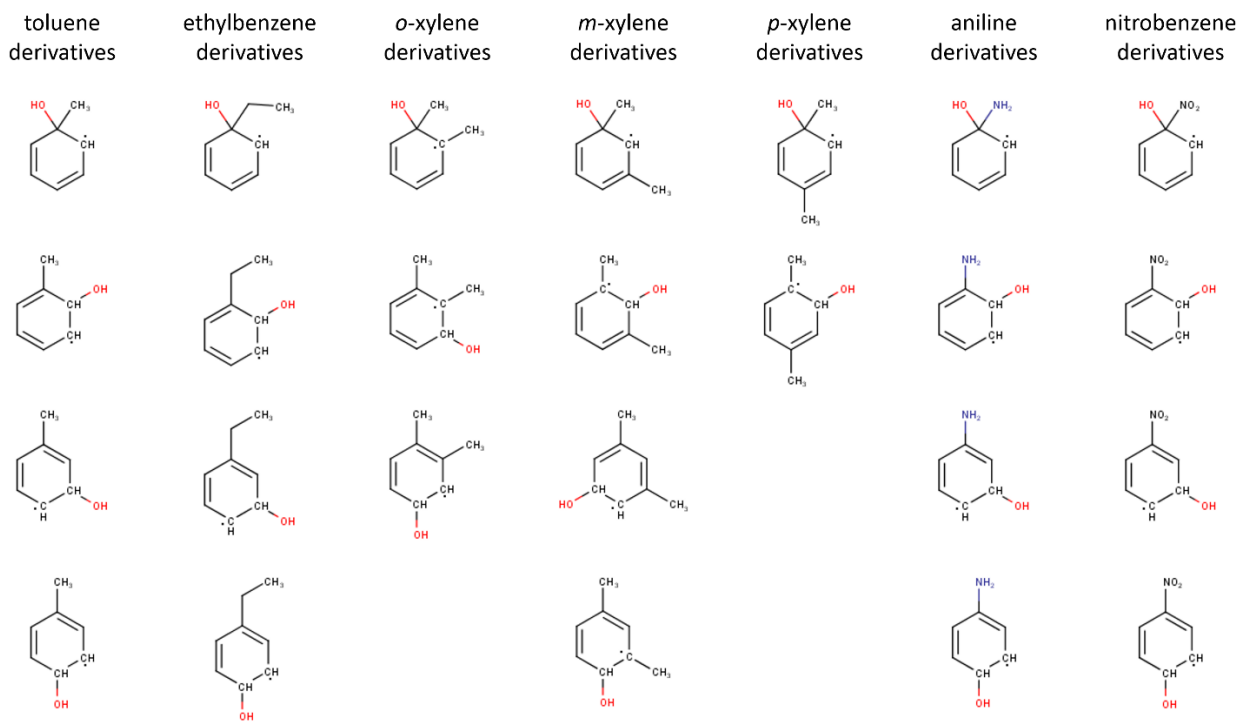

**Figure 7**. All possible products resulting from the reaction of various hydroxylated oxidation products from oxidation of benzene derivatives by sulfate radical and subsequent HO$^-$ addition. The energies of the various products were computed using both DFT and high-level wavefunction based methods to assess the accuracy of all the computational methods used in the main text.

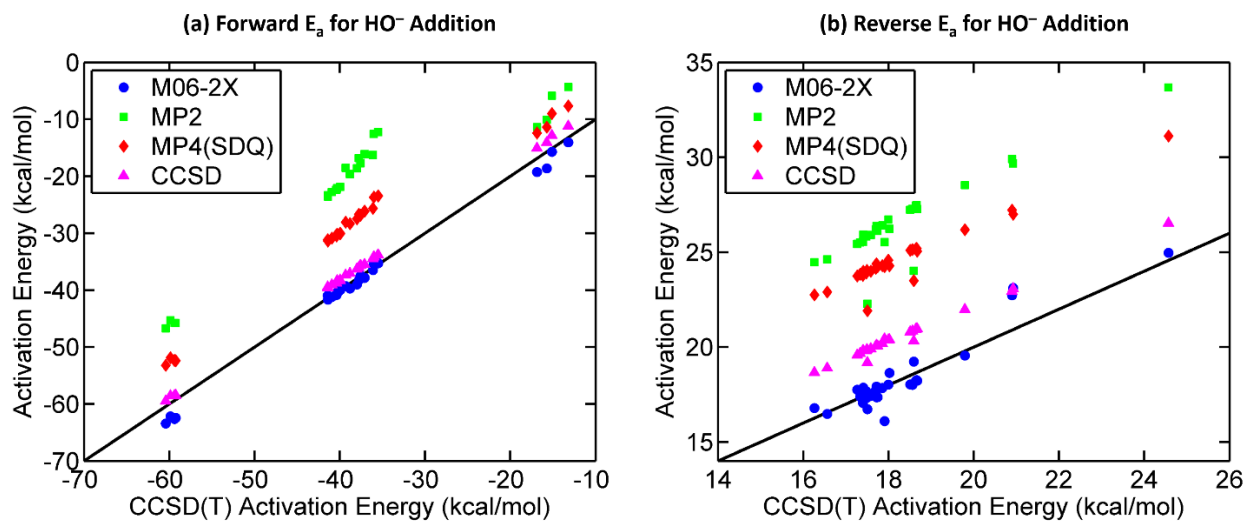

**Figure 8.** Predicted activation energies of the (a) forward and (b) reverse reactions of benzene-derived cation radical reacting with HO⁻. The diagonal line in each figure represents a perfect match to the benchmark CCSD(T) activation energies.

In contrast to the statistical trends for the $SO_4^{\bullet-}$ reaction described previously, we find that the various computational methods more accurately predict the activation energies for the HO⁻ addition, resulting in lower MAEs in general. In particular, both Table 1 and Fig. 8 actually show that the M06-2X functional outperforms all other methods (even the CCSD method) for both the forward and reverse activation energy for the HO⁻ addition. However, we note that the M06-2X $R^2$ values (= 0.88) for the reverse reaction are noticeably worse than the corresponding MP4(SDQ) and CCSD wavefunction-based methods. Consequently, these deviations from $R^2 = 1$ indicate a non-systematic error in the M06-2X activation energies for the reverse reaction for the HO⁻ addition. As before, the MP4(SDQ) and MP2 methods incur larger errors compared to either the CCSD or M06-2X calculations for both the forward and reverse activation energy for the HO⁻ addition. These errors are due to the perturbative nature of the MP4(SDQ) and MP2 methods which do not capture dynamical correlations effects compared to the accurate CCSD(T) calculations. Based on our benchmarks on activation energies alone, we find that the popular M06-2X DFT functional is significantly more accurate for HO$^-$ reactions than for SO$_4^{\bullet-}$ reactions (although further work is needed on assessing the complete reaction pathways). This stark difference in accuracy is due to the training set used to parameterize the M06-2X functional, which primarily consists of hydrocarbon molecules that do not include other non-carbon based environmental species such as SO$_4^{\bullet-}$. As a result, while the M06-2X functional yields impressive (nearly CCSD(T)-quality) accuracy for conventional reactions, we recommend the use of high-level quantum calculations to spot-check environmental chemistry reactions that may lie outside the training set of the M06-2X functional, particularly for water oxidation reactions that involve SO$_4^{\bullet-}$.

**Conclusions**

In conclusion, we have examined a wide set of reactions (110 forward and reverse reactions, in total) that play an important role in sulfate radical-based oxidation processes for water reuse and groundwater remediation. To understand these complex reactions at a fundamental level, we utilized a variety of computational techniques including DFT, MP2/MP4 perturbation theory methods, and high-level coupled cluster CCSD/CCSD(T) approaches. While DFT calculations have started to become commonplace in predicting reaction mechanisms and activation energies in environmental processes, many DFT functionals are highly-parameterized and can fail dramatically for chemical species outside of their training set. As a result, additional high-level methods are essential for spot-checking these results and for iterative feedback between theory and experiment, particularly for accurate calculations of reaction mechanisms in environmental chemistry. Within this comprehensive study, which involves over 100 reactions and 5 different computational methods, we found that the popular M06-2X functional is more accurate for HO$^-$ reactions than for SO$_4^{\bullet-}$ reactions (based on high-level CCSD(T) calculations used as benchmarks). In general, we find that the M06-2X functional does perform reasonably well for both HO$^-$ and $SO_4^{\bullet-}$ reactions; however, we noticed a low degree of statistical correlation for the reverse activation energy barriers in the $HO^-$ reactions. As a result, while the M06-2X functional yields impressive (nearly CCSD(T)-quality) accuracy for conventional reactions, high-level benchmarks should be carried out to spot-check reactions that may lie outside the training set of M06-2X (such as reactions that involve $SO_4^{\bullet-}$ or other inorganic oxidants). These extensive calculations and methodological assessments provide a predictive path towards understanding increasingly more complex reaction mechanisms in environmental processes.

**Acknowledgements**

I.AW. acknowledges the National Science Foundation through the Research Experiences for Undergraduates (REU) program (ACI-1452367). H.L. was supported by the National Science Foundation (CHE-1611306). All DFT and wavefunction-based calculations by S.P. and B.M.W were supported by the U.S. Department of Energy, Office of Science, Early Career Research Program under Award No. DE-SC0016269. We acknowledge the National Science Foundation for the use of supercomputing resources through the Extreme Science and Engineering Discovery Environment (XSEDE), Project No. TG- ENG160024. The six reviewers are acknowledged for their helpful and constructive comments.